\documentclass[%
 reprint,
 amsmath,amssymb,
 aps,
]{revtex4-2} 

\usepackage{graphicx}
\usepackage{dcolumn}
\usepackage{bm}
\usepackage{amsfonts}
\usepackage{amsmath}
\usepackage{amssymb}
\usepackage{float}
\usepackage{caption}
\usepackage{color}
\usepackage{cleveref}

\usepackage{graphicx,subcaption}

\begin{document}

\preprint{APS/123-QED}

\title{Effects due to generation of negative frequencies during temporal diffraction}

\author{E. Hendry$^{1*}$, C. M. Hooper$^1$, W. P. Wardley$^1$, and S. A. R. Horsley$^1$
\\$^1$School of Physics and Astronomy, University of Exeter, Stocker Road, Exeter, EX4 4QL, UK
\\$^*$Corresponding author: e.hendry@exeter.ac.uk
}

%
%
\begin{abstract}
Temporal diffraction from rapidly time-modulated materials can generate new frequency components not present in an incident wave.  In such an experiment, the spectral extent of these new frequencies is determined by the rate of modulation relative to the period of the oscillating field. Here we present a temporal diffraction experiment carried out in the far-infrared (THz) spectral region. Using graphene, a fast modulator for this spectral domain, we show that one can modulate the transmission significantly faster than the period of a narrow band THz field. This leads to a large bandwidth for the generated frequencies, including the generation of negative frequency components. We show that interference between negative and positive frequency components give rise to distinctive oscillatory features in the transmitted spectrum.
\end{abstract}

\maketitle


%
%
\section{\label{Sec: Introduction}Introduction}

%
%
\begin{figure}
    \centering
    \includegraphics[width=0.65\linewidth]{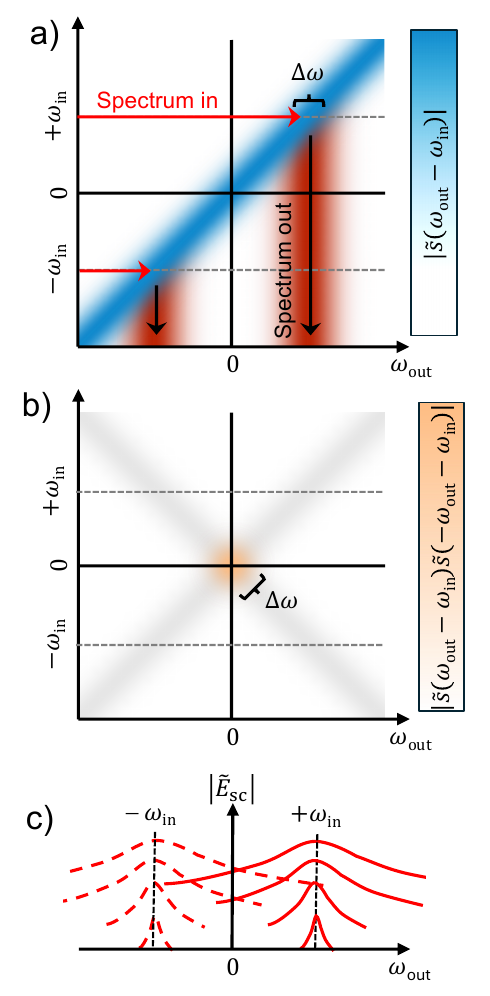}
    \caption{
    \textbf{Frequency scattering due to time-varying media:}
    (a) The scattering amplitude, $\widetilde{s}\left( \omega_{\rm out}-\omega_{\rm in}\right)$, (shaded in blue) tells us the wave amplitude at frequency $\omega_{\rm out}$, due to an incident wave at frequency \(\omega_{\rm in}\).  Here this is centred around $\omega_{\rm out}=\omega_{\rm in}$, with the characteristic width \(\Delta \omega\) increasing with the rapidity of the modulation.  We sketch a monochromatic excitation at \(\pm\omega_{\rm in}\) (grey horizontal lines) leading to a broadband scattered spectrum (shaded in red).
   (b) The coupling between positive and negative frequencies in the scattered intensity is given in the final term of Eq. (\ref{eq:interference}) and is proportional to the product, $\left|\widetilde{s}\left( \omega_{\rm out}-\omega_{\rm in}\right)\widetilde{s}\left( - \omega_{\rm out}-\omega_{\rm in} \right)\right|$ (shown in orange).  We can obtain this product by taking the blue region in panel a and multiplying it with its copy, flipped across the $x$--axis  (shown in grey).  In this example, the input frequencies $\pm\omega_{\rm in}$ lie outside the orange region, and thus negligible phase-sensitivity is expected.
    (c) The scattered spectrum broadens as the modulation rate increases, here indicated as vertically offset copies of the output spectrum. If the modulation rate is fast enough (\(\Delta \omega \gtrsim \omega_{\rm in}\)), positive and negative frequency components of the scattered field will interfere.
    }
    \label{fig: ScatteringMatrices}
\end{figure}

Despite running counter to our intuition of what ``frequency'' means, both positive \emph{and} negative frequencies are required to decompose any time domain signal $s(t)$ into its Fourier components, $\tilde{s}(\omega)$.  Yet these Fourier components inevitably have the symmetry $\tilde{s}(\omega)=\tilde{s}^{\ast}(-\omega)$.  This means any negative frequency amplitude can be inferred from the corresponding positive frequency one, allowing us to ignore the negative frequency part of a spectrum, a perspective usually taken when calculating wave propagation.

This perspective is less clear in systems that exhibit rapid motion or time modulation~\cite{galiffi2022}, where a large enough frequency shift may occur to invert the sign of the wave frequency.  Such an inversion implies a time-reversal of the wave, leading to a range of physical effects.  For example, for the Doppler effect, Lord Rayleigh first showed that motion faster than the phase velocity of a wave leads to a transformed frequency that has the opposite sign to the laboratory frequency, becoming its negative for motion at twice the wave speed~\cite{rayleigh2018theory}.  This sign change implies that the moving observer will see the laboratory frame signal played in reverse.  Such large Doppler shifts have recently been observed for rotational motion~\cite{gibson2018}, where the change in the sign of the frequency leads to the Zeldovich effect~\cite{zeldovich1971}, a time-reversal of material absorption to give amplification~\cite{cromb2020,braidotti2024}.  Cherenkov radiation~\cite{volume8}, the transformation of a static field into propagating radiation due to motion, also occurs when a charge moves faster than the phase velocity of light in a medium, with the cone of radiation defined by the condition for the frequency to change sign between laboratory and rest frame~\cite{svidzinsky2021}.  In the same vein, quantum friction~\cite{pendry1997} - the lateral Casimir force between sliding plates - arises due to the absorption of surface waves whose motion has been time-reversed due to a large Doppler shift~\cite{horsley2012}.  These phenomena are also connected to the Hawking effect, which is well understood to be due to the mixing of positive and negative frequency waves at the event horizon, analogues of which can be found in both fluid dynamics and optics~\cite{rousseaux2008}.

Although the above effects are due to relative motion (i.e. the negative frequencies arising from the Doppler shift) there has been recent interest in negative frequencies arising due to rapid time modulation of material parameters~\cite{galiffi2022}.  Perhaps the most well-known effect of negative frequencies in this context is parametric amplification~\cite{brunner1977}, where a small modulation of the refractive index at twice the wave frequency couples the wave and its time reverses, leading to a standing wave of ever growing amplitude.  A lot of current work is concerned with high contrast, ultrafast modulations of material parameters, where the temporal analogue of reflection occurs, splitting the wave into counter-propagating waves of positive and negative frequency. This effect has been observed for both water waves~\cite{bacot2016}, and more recently for radio frequencies~\cite{moussa2023,Galiffi2023}, allowing a wave-field to be time reversed in response to the application of a short modulation pulse.  This promises more extreme amplification phenomena ~\cite{Feinberg2025}, including photon pair creation, and a combination of space and time modulation can lead to novel traveling wave amplifiers~\cite{galiffi2019,horsley2023,horsley2024}.  

For frequencies higher than radio-frequencies, rapid, high-contrast modulations of the refractive index through optical electron heating in epsilon near zero materials have recently been shown to produce strikingly different electrodynamics from passive materials.  In some of the first demonstrations of this concept in Refs.~\cite{Zhou2020,Khurgin2020,bohn2021b,tirole2022}, temporal diffraction is generated using ITO modulators working at near-infrared frequencies, where the modulation time (the time taken to switch the optical transmission or reflection) is considerably slower than the characteristic period of the electromagnetic wave. For example, when ITO is excited by a $\sim$100fs laser pulse, the modulation time is around 5\% of the period of the radiation ~\cite{bohn2021a}, which can generate frequency components just outside the bandwidth of the near IR pulses used in experiments~\cite{bohn2021b,tirole2022}. However, because of the relatively low modulation rate, the frequency shifts induced by temporal diffraction using this approach are typically far from time-reversing the incident wave. While it is difficult to modulate materials with high efficiency significantly faster than this for near-IR frequencies, relatively faster modulation can be achieved for longer infrared wavelengths.

In this paper, we demonstrate that an abrupt temporal modulation of the refractive index of graphene can be significantly faster than the field oscillation period in the far-infrared. For a 0.5 THz narrow band field, we show that a modulation rate of $>$1000\% of the radiation frequency can be achieved, such that an incident far-infrared field can be subject to large enough frequency shifts to time reverse part of the incident signal. This conversion of positive to negative frequencies gives rise to phase sensitivity of the transmitted intensity, which we observe directly in experiment.

To gain a simple understanding of this effect, consider temporal diffraction of a monochromatic field $E_{\rm in}$ with an angular frequency \(\omega_{\rm in}\) and complex amplitude $E_0$,
\begin{equation}
    E_{\rm in}(t) = {\rm Re}\left[{\rm e}^{- \mathrm{i}\omega_{\rm in}t}E_0\right].
    \label{eq:incident_fields}
\end{equation}
When incident upon a time-varying material, temporal diffraction leads to a spread in frequency while momentum remains constant, analogous to how spatial diffraction leads to a spread in momentum while frequency remains constant. Ignoring dispersion, the scattered field \(E_{\rm sc}(t)\) is given by the product of the incident field \(E_{\rm in}(t)\) and an instantaneous scattering coefficient \(s(t)\), i.e. $E_{\rm sc}(t)=s(t)E_{\rm in}(t)$.  The corresponding Fourier spectrum of the scattered field can then be written in terms of the Fourier transform of the scattering coefficients, \(\widetilde{s}(\omega_{\rm out})\). Evaluating this, and noting that \(s(t)\) is required to be real, one finds
\begin{align}
    {\widetilde{E}}_{\rm sc}\left( \omega_{\rm out}\right) &= \frac{1}{2}\left[\widetilde{s}\left( \omega_{\rm out} - \omega_{\rm in} \right)E_{0} + \widetilde{s}^{\ast}\left(-\omega_{\rm out} - \omega_{\rm in} \right)E_{0}^{\ast}\right]\nonumber\\
    &=\frac{1}{2}\left[\widetilde{s}_{+}E_0+\widetilde{s}_{-}^{\ast}E_0^{\ast}\right].
    \label{eq:scattered_spectrum}
\end{align}
In the absence of time modulation, $\widetilde{s}$ is proportional to a delta function, and the scattered amplitude is zero except when $\omega_{\rm out}=\pm\omega_{\rm in}$, i.e. frequency is conserved between incident and scattered fields.  Meanwhile, time-variations in the material generally give rise to a polychromatic scattered spectrum from a monochromatic input.  The bandwidth of this scattered spectrum depends on modulation rate.  If the properties of a material change over a characteristic timescale \(\tau\), the Fourier spectrum of the scattered field is primarily confined to the region \(|\Delta\omega| = \frac{2\pi}{\tau}\), as indicated by the diagonal blue region in Fig.~\ref{fig: ScatteringMatrices}(a), where we sketch the spreading of a monochromatic input due to the time modulation of the scattering amplitude, the resulting output spectrum represented in red.

For a modulation timescale much shorter than the oscillation period of the incident wave, the output spectrum will be spread such that a significant portion crosses the origin, mixing positive and negative frequency input amplitudes (the two red regions in Fig.~\ref{fig: ScatteringMatrices}(a) would then overlap).  Just as for the interference between waves emerging from two slits, this overlap in Fourier space exhibits interference and a consequent sensitivity to the relative phase between the positive and negative frequency Fourier amplitudes.  However, unlike the interference between two spatially separated slits, here the interference indicates an oscillation of the scattered \emph{energy} with respect to phase, indicating potential amplification.  This relative phase between Fourier components can be simply adjusted through introducing a time-delay, \(t_{0}\), between the modulation of the material and the arrival time of the probe, $E_0={\rm e}^{\mathrm{i}\omega_{\rm in}t_0}\left|E_0\right|$.  The scattered intensity at a given frequency $\omega_{\rm out}$ is then given by
\begin{equation}
    \left| {\widetilde{E}}_{\rm sc} \right|^{2} = \frac{\left|E_{0}\right|^2}{4}\left(\left|\widetilde{s}_{+}\right|^{2} + \left|\widetilde{s}_{-}\right|^{2} +2\,{\rm Re}\left[{\rm e}^{2\mathrm{i}\omega_{\rm in}t_{0}}\widetilde{s}_{+}\widetilde{s}_{-}\right]\right)\label{eq:interference}
\end{equation}
which equals the sum of the intensities centred around positive and negative frequency, $|\widetilde{s}_{\pm}|^2$, plus the interference between the two, $2\,{\rm Re}\left[{\rm e}^{2\mathrm{i}\omega_{\rm in}t_{0}}\widetilde{s}_{+}\widetilde{s}_{-}\right]$, which oscillates with the arrival time of the incident field.    In slowly varying media, the functions \(\widetilde{s}_{\pm}\) will be confined almost entirely to the positive/negative half of the frequency axis.  The overlap, $\widetilde{s}_{+}\widetilde{s}_{-}$ thus vanishes, and the scattered intensity becomes independent of the time-shift \(t_{0}\). By contrast, in rapidly varying media, positive to negative frequency conversion allows the functions \(\widetilde{s}_{\pm}\) to overlap.  Where this occurs the system becomes sensitive to shifts of the incident field in time, with a characteristic frequency of \(2\omega_{{\rm in}}\). Indeed, oscillations of this form are characteristic of positive to negative frequency conversion in any time-varying system, occurring when a material changes with a timescale \(\tau\) sufficiently fast that \(\frac{2\pi}{\tau} > \omega_{\rm in}\).

%
%
\section{\label{Sec: Experiment}Experiment}
For rapid modulation in the far-infrared, graphene is the material of choice~\cite{Zhou2023}. The high electron mobility in graphene gives rise to strong conductivity response at low frequency, such that a single layer of CVD grown graphene can modulate transmission of far infrared radiation by more than $\sim$10\%~\cite{Tomadin2018} (sufficient for this study), and can be considerably higher than this if incorporated into resonator structures ~\cite{Gopalan2019, Kim2023}. Meanwhile, the linear band structure means that scattering is remarkably efficient, giving rise to very short thermalization times~\cite{Tomadin2018}. The photophysics of graphene when photoexcited by a femtosecond optical pulse is complex, with a competition between interband and intraband heating effects. The largest modulation of the transmission of a single graphene layer is normally observed for intraband heating, which dominates when the Fermi level is more than 150meV from the Dirac point~\cite{Tomadin2018}. For these high doping levels, the intraband heating following photoexcitation by an optical femtosecond pulse leads to a shift of the fermi level due to the energy dependent density of states, a lower conductivity, and a photo-induced step increase in transmission. 

\begin{figure}
    \centering
    \includegraphics[width=0.8\linewidth]{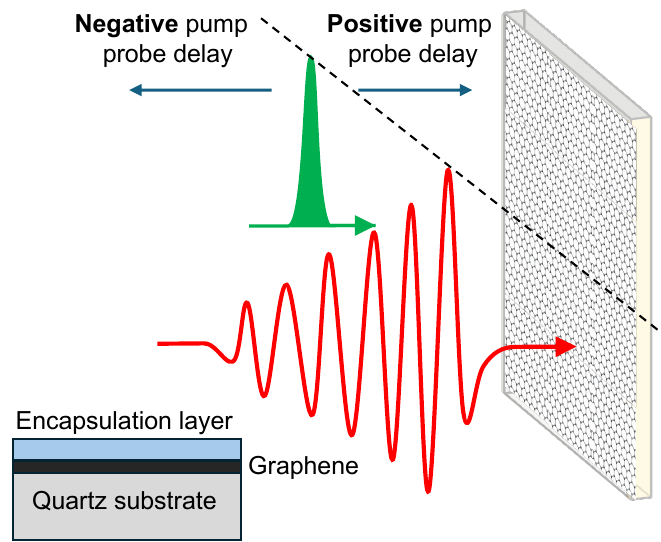}
    \caption{\textbf{Experimental set--up}:  Schematic defining negative and positive pump – probe delay. Dotted line indicates time zero. Inset depicts the sample cross-section (dimensions given in text).}
    \label{fig:experiment}
\end{figure}

For CVD-grown graphene on a substrate, it is natural for such high doping levels to exist due to adsorbed water molecules.  However, with the light intensities used in our experiments here, we observe that the Fermi level of bare graphene gradually decreases with increasing exposure to femtosecond laser pulses. To retain doping under illumination, monolayer graphene on a quartz substrate (Graphene Supermarket) is encapsulated by sequentially spin-coating two 200nm layers of Poly(methyl methacrylate) (950K A4, Kayaku Advanced Materials) giving the structure depicted in Fig.~\ref{fig:experiment}. This encapsulation ensures stability of the sample even for the highest intensities used here.  As we see below in Fig.~\ref{fig:pulses}, the characteristic width of the step function in transmission induced on photoexcitation is only slightly broader than our femtosecond laser pulses, i.e. significantly faster than the period of a 0.5 THz field.

We use a THz time domain measurement adapted from that described in~\cite{Hornet2016}. Briefly, an amplified femtosecond laser system (800 nm, 1 kHz repetition rate, $\sim$80 fs) to generate and detect our THz probe beam in a pair of ZnTe crystals through optical rectification and balanced electro-optic sampling, respectively. This allows us to determine the electric field, $E$, of a single cycle THz pulse (central frequency $\sim$0.8 THz and width $\sim$1.0 THz). For the experiments presented below, we pass this THz pulse through a narrow band pass filter (0.5 THz central frequency, Swiss THz), and measure the narrow band pulse transmitted through our sample, shown in Fig.~\ref{fig:pulses}(a) and (b).  A narrow band of incident frequencies is critical here to be able to observe the frequencies of the temporally scattered field generated outwith this bandwidth.
%
%
%
%
\begin{figure*}
    \centering
    \includegraphics[width=0.8\linewidth]{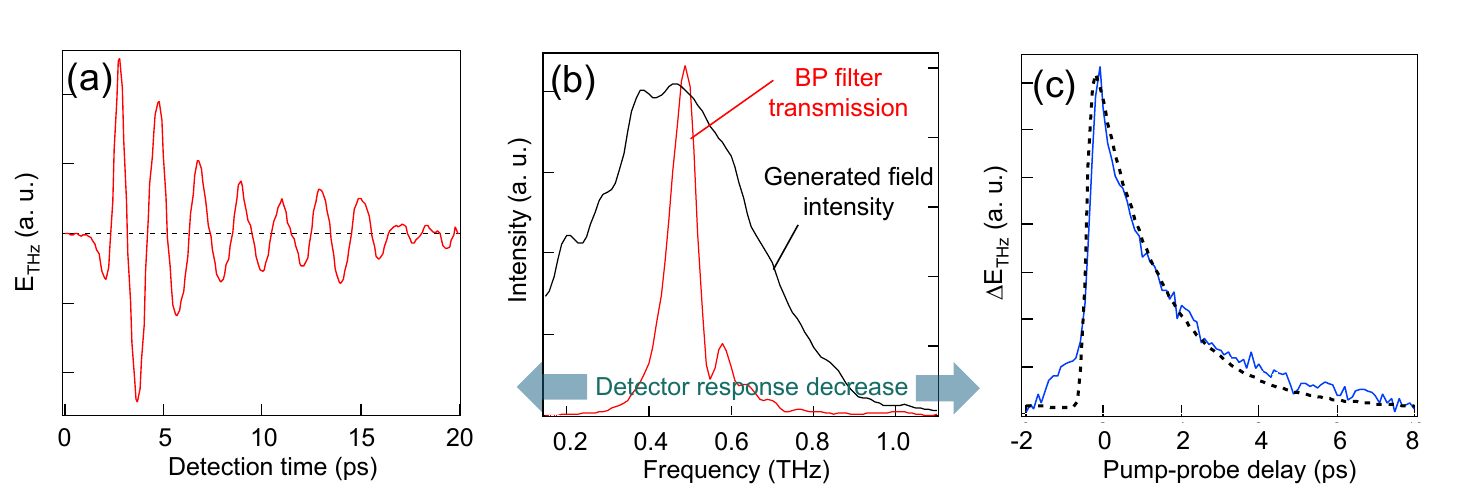}
    \caption{\textbf{Incident and transmitted probe pulse:}  (a) Time domain measurement of field transmitted through narrow band pass filter centred at 0.5 THz. (b) Red: Spectrum of the pulse transmitted through band pass filter. Black: Broad spectrum of modulated field introduced by ultrafast excitation of the sample, limited in spectral range by the detector response function. The spectrum for the generated intensity is expected to be even broader than measured due to the frequency dependence of the detector response function~\cite{Hendry2004}. (c) The temporal modulation of transmission observed when measuring at the peak of the THz probe field and varying the pump-probe delay.}
    \label{fig:pulses}
\end{figure*}

The femtosecond laser provides a pump pulse (fluence 3 J/m\textsuperscript{2}) to modulate the graphene. The temporally scattered field, $\Delta E_{\text{THz}}$, is defined as

\begin{equation}
    \Delta E_{\text{THz}} = E_{\text{THz}}^{\text{(pump-on)}} - E_{\text{THz}}^{\text{(pump-off)}}.
    \label{eq:energy_difference}
\end{equation}

\noindent As we vary the time between the optical pump and THz probe pulses, we record the dynamics shown in Fig~\ref{fig:pulses}(c). The photoexcitation pulse arrives at $\sim$0 ps, defined by the temporal overlap between the pump pulse and the peak THz field, with the sign convention used for the time delay illustrated in Fig.~\ref{fig:experiment}. Note that this measurement at the peak THz field represents a section of the full 2D dataset (shown as vertical dotted line shown in Fig.~\ref{fig:results}(a). In Fig.~\ref{fig:experiment}(c), we observe a very fast rise in  transmitted field associated with rapid heating and thermalization of electrons, peaking at an increase in transmission of around 12\%, followed by picosecond relaxation times associated with carrier cooling, similar to that observed previously~\cite{Tomadin2018}. The dotted line is a fit of the temporal response described by the convolution $\exp{(-t^2/\tau_{\text{rise}}^2)} * \exp{(-t/\tau_{\text{decay}})}$, with $\tau_{\text{rise}}$=160 fs and $\tau_{\text{decay}}$=1.7 ps. While the amplitude of the step increase in transmission changes with pump intensity (saturating for fluences above 5 J/m\textsuperscript{2}) the temporal dynamics are not strongly sensitive. We therefore use this convolution as characteristic of the temporal response function of the sample.

%
%
\section{\label{Sec: Results}Results and Discussion}
\begin{figure*}
     \centering
    \includegraphics[width=1\linewidth]{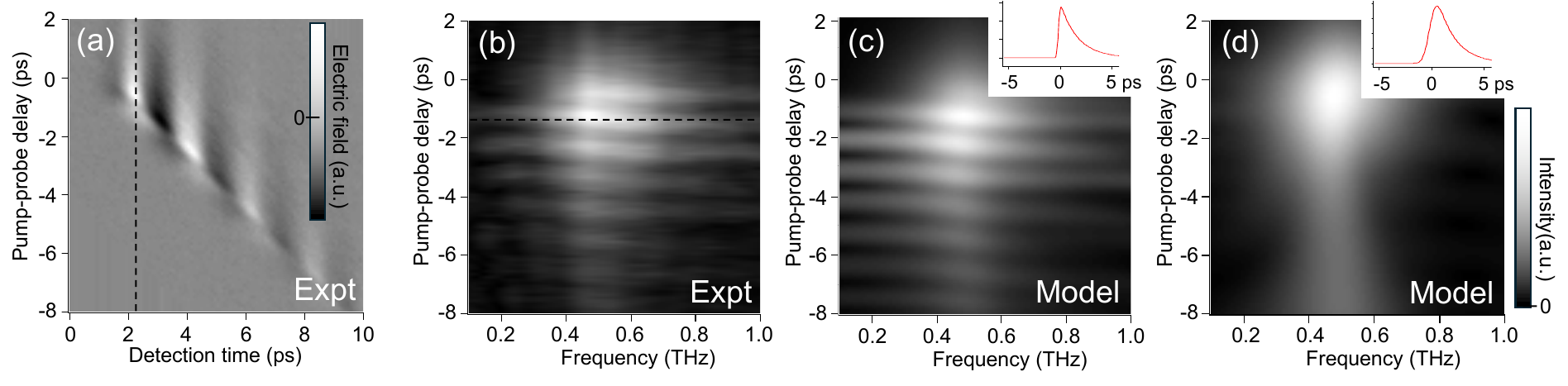}
    \caption{\textbf{Observed positive/negative frequency coupling:}  (a) Experimental results showing the modulation of the transmitted THz field as a function of detection time delay and pump-probe delay. The dotted line indicates the cross section of data shown in Fig.~\ref{fig:pulses}c. (b) Experimental spectra resulting from Fourier transform of the detection time axis in (a). The dotted line indicates the cross section plotted as the black line in Fig.\ref{fig:pulses}b. (c) Modeled spectra resulting from a  convolution between the measured incident field and the temporal response function of the graphene. Inset shows the temporal response function used. (d) Modeled spectra when the rise time of of the temporal response function is assumed to be slower than measured, with $\tau_{\text{rise}}=500$ fs.}
    \label{fig:results}
\end{figure*}

In Fig.~\ref{fig:results}(a), we plot $\Delta E_{\text{THz}}$ as a function of the pump--probe delay and detection time delay. Due to the very fast response of the graphene, each individual peak and trough of the probe field is distinct in the scattered field. The sharp onset to the scattered field is crucial for generating the broad range of frequencies necessary for interference between positive and negative frequency components. When we Fourier transform the detection time delay axis, we get the spectra presented in Fig.~\ref{fig:results}(b). The scattered intensity due to the modulation is very broad, covering the entire spectral range shown, only falling to zero for frequencies beyond 1 THz. We take a cross section of this data represented by the dotted line in Fig.~\ref{fig:results}(b), and compare it to the incident intensity in Fig.~\ref{fig:pulses}(b). We see that the spectral composition of the scattered intensity is much broader than the incident intensity. Indeed, we expect the true composition of the scattered intensity to be even broader than measured, as both the low and high frequency components are not efficiently detected by our THz detection scheme~\cite{Hendry2004}.

In Fig.~\ref{fig:results}(c) we plot the spectra predicted from a convolution between the measured incident field and the temporal response function of the sample. We see that there is good agreement between this and the measured spectra in Fig.~\ref{fig:results}(b). Most striking in both the experimental results and model is the characteristic oscillation of the scattered spectrum as a function of the pump-probe time which covers the entire spectrum. Curiously, the minima in the oscillating scattered field do not coincide with the roots of the incident field - see supplementary information. As described in the introduction, this oscillitory behavior is precisely the signature expected for such rapid modulation, the calling card of positive to negative frequency conversion. 

Note that this oscillatory quality requires the rapid change to transmission shown in Fig.~\ref{fig:pulses}(c). If the dynamic change is significantly slower (for our incident frequency of 0.5 THz characterized by $\tau_{\text{rise}}\ge$500 fs) we enter the regime more comparable to recent ultrafast modulation of THz metamaterials~\cite{Tunesi2021} and the ITO temporal diffraction experiments in the near-IR~\cite{Zhou2020,Khurgin2020,bohn2021b,tirole2022}. When we model the expected result from slower dynamics in Fig.~\ref{fig:results}(d), we still observe a scattered spectrum that is significantly broader than the incident intensity, but without the oscillatory behavior that is characteristic of positive to negative frequency conversion.  This highlights the additional interference effects that occur when the material parameters can be varied on a sub--wave period timescale.

%
%

Finally, in the both the experimental and modeled spectra in Figs.~\ref{fig:results}(b) and (c), we observe an interesting phenomenon where the phase of the oscillation with pump-probe delay changes by almost $\pi$ as the frequency of the scattered field moves through the incident frequency at 0.5 THz. This shift is due to the causal, and thus inherently asymmetric, temporal response of the sample to the pump pulse. Indeed, this is well-modeled by assuming a dispersionless transmission coefficient, rising as a step and decaying as $\exp(-t/\tau_{\rm decay})$  The Fourier transform of this is ${\rm i}[\omega-\omega_{\rm in}+{\rm i}/\tau_{\rm decay}]^{-1}$, which implies a total phase shift of $2 \arg(\tau_{\rm decay}^{-1}+\mathrm{i} \omega_{\rm in}) = 2.8\,\mathrm{rad}$ over the frequency window $0$ to $1$ THz.

%
%
\section{\label{Sec: Conclusions}Conclusions}
We present a temporal diffraction experiment carried out in the far-infrared spectral region. Using a fast modulator, graphene, we show that one can increase transmission significantly faster than the THz field period, leading to a large bandwidth for the generated frequencies, including the generation of negative frequency components. We show that these negative frequency components give rise to distinctive oscillatory features in the transmitted spectrum due to interference. The observation of positive to negative frequency conversion in our experiments suggests that graphene might be an interesting platform to observe extreme amplification phenomena in the far infrared induced by temporal modulation - an intriguing prospect for a spectral region which suffers from a lack of conventional, high power radiation sources.

\bibliography{refs} 

\end{document}


\begin{Large}
\noindent{\textbf{Supplementary Information}}
\end{Large}

\section{Relation of minima in scattered amplitude to the roots of incoming field}

To understand the spectral location of the maxima and minima in the transmitted field relative to the incoming field, we need to consider the interaction between the pump and terahertz pulse, particularly the implications of a pulse having a smaller width than the cycle time of the probe wave. If we consider the pump as a Gaussian pulse with a shorter pulse width than the terahertz wave period, ($\Delta\ll2\pi/\omega_c$), then being shorter than the cycle must mean, once Fourier transformed, the transmitted field has a bandwidth greater than the central frequency. That means $\omega_c \cdot \Delta $ must be smaller than 1 and, with the Gaussian centred at the Terahertz probe's carrier frequency, $\omega_c$, this results in a spectrum that spreads across the origin of the frequency axis, coupling positive and negative frequencies.

\begin{figure}
    \centering
    \includegraphics{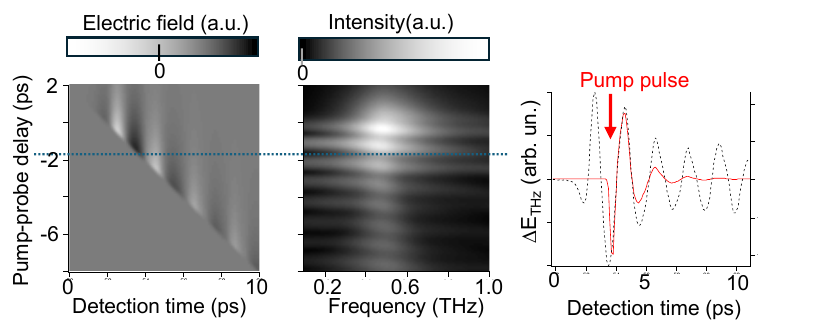}
    \caption{
    \textbf{Figure S1 The left hand panel in the figure below shows the calculated scattered field as a function of pump-probe delay and detection time, for the assumed material response function as defined in the main text. The middle panel shows the corresponding intensity spectrum. The blue dotted lines indicate the pump-probe delay for which a minimum is observed for the lowest frequencies in the spectrum. The red line in the right hand panel shows the cross section of the temporally scattered field for the pump-probe delay corresponding to the blue dotted line. The black dotted line in the top right graph shows the incident field.}}
     \label{eq:fieldzeros}
    \end{figure}

In this circumstance, we need to consider the phase of the Fourier components of the pulse.  When the positive and negative frequency Fourier components overlap and are in phase, we will see maximal constructive interference and hence a greater overall transmitted pulse energy.  Conversely, when they are out of phase we have destructive interference and hence a reduced transmission.  To quantitatively illustrate this point, we consider the Fourier spectrum $\tilde{s}$ of a transmitted pulse of carrier frequency $\omega_c$ (which represents a single rapid modulation of our transmitted signal), 
\begin{equation}
  \tilde{s} (\omega) = \sqrt{\frac{\pi \Delta^2}{2}} \left[ {\text{e}^{- \text{i}\phi}}  \text{e}^{-\frac{\Delta^2 (\omega_c - \omega)^2}{2}} +{\text{e}^{\text{i} \phi}}  \text{e}^{- \frac{\Delta^2 (\omega_c +\omega)^2}{2}} \right]\label{eq:fourier-amplitude},
\end{equation}
where the phase $\phi$ represents the offset between the carrier oscillations due to the terahertz probe field and the incoming Gaussian pump pulse.  The total energy in the pulse is proportional to the time integral of $[s(t)]^2$ (i.e. the integral of the Poynting vector), which---via Parseval's theorem---can be computed from the frequency integral,
\begin{align}
    \text{Total energy}&\propto\int_{-\infty}^{\infty}\frac{{\rm d}\omega}{2\pi}|s(\omega)|^2\nonumber\\[5pt]
    &=\int_{-\infty}^{\infty}\frac{{\rm d}\omega}{2\pi}\frac{\pi\Delta^2}{2}\left[{\rm e}^{-\Delta^2(\omega_c-\omega)^2}+{\rm e}^{-\Delta^2(\omega_c+\omega)^2}+2\cos(\phi){\rm e}^{-\Delta^2(\omega^2+\omega_c^2)}\right]\nonumber\\[5pt]
    &=\frac{1}{2}\sqrt{\pi\Delta^2}\left[1+\cos(2\phi){\rm e}^{-\Delta^2\omega_c^2}\right].\label{eq:frequency-domain-integral}
\end{align}
Importantly for this current discussion, the energy includes a phase dependent term, proportional to $\cos(2\phi)$. This phase dependence represents the interference between positive and negative frequencies.  For us $\phi$ is fixed by the relative delay between the pump and probe pulses and the $\phi$ dependent oscillation evident in Eq. (\ref{eq:frequency-domain-integral}) is seen in the data given in Fig. \ref{eq:fieldzeros}, both in terms of the frequencies generated but also in terms of their spectral location.

There is an important observation to make here.  In Eq. (\ref{eq:frequency-domain-integral}) we can see that a minimum in the transmitted energy occurs when $\phi$ is an odd multiple of $\pi/2$.  This is precisely when the carrier oscillation passes through zero just as the pump pulse comes to its peak.  In general however, there is no such association.  Taking a general pump probe profile $p(t)$, we can write its Fourier transform as
\begin{align}
    \tilde{p}(\omega)&=\int_{-\infty}^{\infty}{\rm d}t\,p(t){\rm e}^{{\rm i}\omega t}\nonumber\\[10pt]
    &=\int_{-\infty}^{\infty}{\rm d}t\,p(t)\,\cos(\omega t)+{\rm i}\int_{-\infty}^{\infty}{\rm d}t\,p(t)\,\sin(\omega t)\nonumber\\[10pt]
    &=|\tilde{p}(\omega)|{\rm e}^{{\rm i}\psi(\omega)}
\end{align}
It is only for a time symmetric pump profile that $\tilde{p}$ is real, at which point the phase of the Fourier amplitude, $\psi$ must be a multiple of $\pi$.  However, for a temporally asymmetric pump $\psi$ can take arbitrary values.  Using such a pulse in Eq. (\ref{eq:frequency-domain-integral}) we can see that this phase will enter the cosine, making the replacement $\cos(2\phi)\to\cos(2(\phi-\psi))$, thus displacing the temporal location of the minima in the transmitted signal.

Figure S1 demonstrates the relation expected for the experimentally measured probe field and modulation dynamics in graphene, calculated using the convolution model described in the main text. If we consider a pump probe delay for which we observe minimal scattered intensity at low frequencies (indicated by the blue dotted line),  we can compare to the incident field (black dotted line in right-hand panel of S1) the time-domain scattered field by extracting the cross section along the blue line (red line in right-hand panel of S1). We see that, somewhat counter-intuitively, the minimum in the scattered spectrum occurs when the pump pulse coincides in time with a \textit{maximum} in the incident field. As outlined above, this is a property which results from the assymetric nature of the temporal modulation.

    \section{Importance of fast rise time of modulation in generating negative frequencies}

A key factor in the generation of negative frequencies is the speed at which a material is modulated, something we can demonstrate by simulating a range of rise times and looking at the generated spectrum.

In Figure S2 we plot the simulated scattered THz
field as a function of detection time delay and pump-probe delay for a range of pump pulse rise times. The oscillatory features in the spectra in the middle column of the figure occur due to the the intereference of positive and negative scattered frequency components, as described in the main text. The scattered field for $\tau_{rise}$= 150fs (top right panel of the figure) has a very sharp onset followed by a damped oscillation – it is this functional form for generating the interference condition. As we slow the rise time of the modulation function, the banding effects in the spectra both fade in their intensity, as seen in their reduced contrast, but also in their width. This results, for the slowest rise time, in a scattered field with much lower frequency broadening and none of the banded interference effects. Moreover, as $\tau_{rise}$ increases, we see the evolution of the functional form of the scattered field in the right-hand column into a more oscillatory functional form, which have Fourier components localized around the central incident frequency. 

  \begin{figure}
    \centering
    \includegraphics[width=0.65\linewidth]{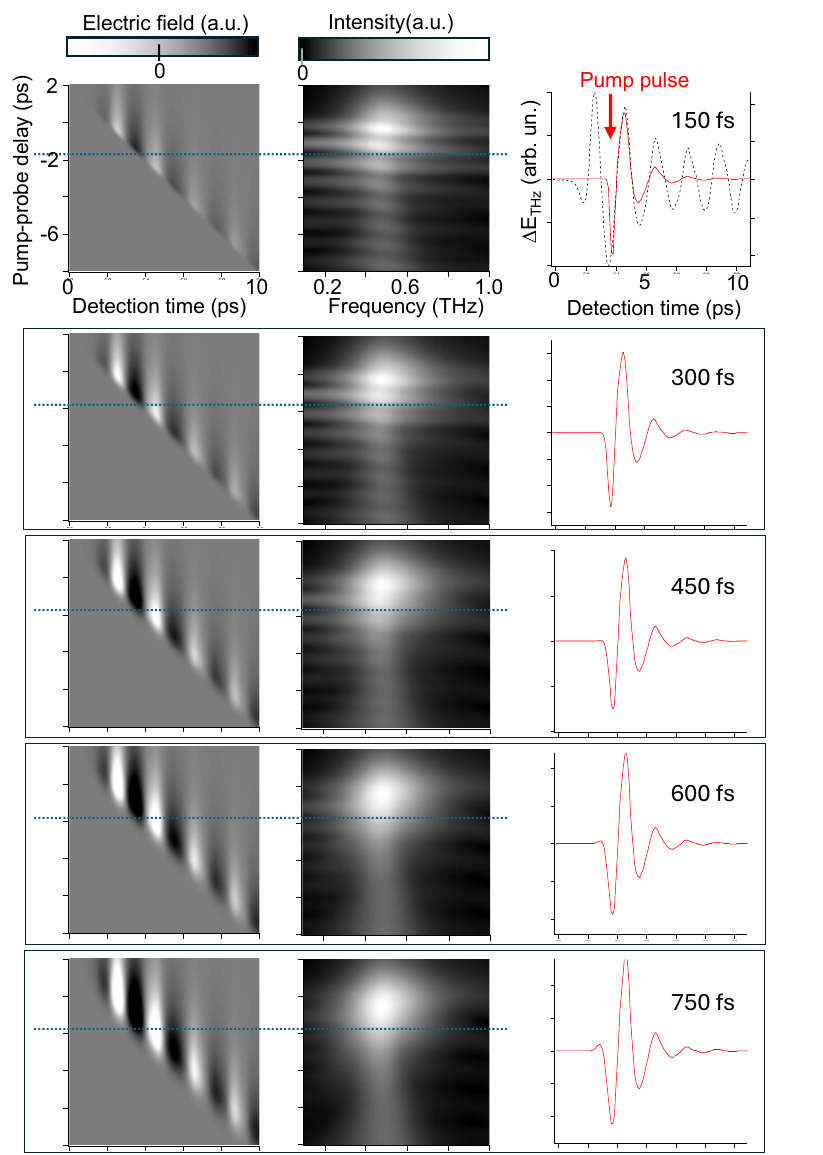}
    \caption{
    \textbf{Figure S2 Each additional row shows time domain scattered field (left), intensity spectrum (center) and field cross section across the blue dotted line, each with increasing rise time. One observes the effects arising from negative frequencies fading as the rise time increases}}
   
     \label{eq:varyrisetime}
  \end{figure}